\documentstyle[11pt,newpasp,twoside,epsf]{article}
\def\edcomment#1{\iffalse\marginpar{\raggedright\sl#1\/}\else\relax\fi}
\reversemarginpar

\begin{document}
\title{Metal Abundances and Kinematics of the Ly-$\alpha$ absorbers}
\author{Sergei A. Levshakov}
\affil{Department of Theoretical Astrophysics, Ioffe Physico-Technical Institute, 
Politekhnicheskaya Str. 26,
194051 St. Petersburg, Russia}

\begin{abstract}
Both high resolution spectra of QSOs observed at the
8-10~m telescopes and advanced methods of data analysis are crucial 
for accurate measurements
of the chemical composition and physical parameters of
the intervening clouds. 
An overview of our recent results obtained with
the Monte Carlo inversion (MCI) procedure
is presented.
This includes: (1) variations of the shape of the local background ionizing
continuum in the 1--5~Ryd range at redshift $z \sim 2.8-3.0$; 
(2) an inverse correlation between the measured metallicity, [C/H], and the absorber
line-of-sight linear size, $L$;
(3) a functional dependence between the line-of-sight velocity dispersion,
$\sigma_v$, and $L$. 
\end{abstract}

\section{Introduction}

A study of quasar (QSO) absorption-line spectra is generally recognized as the most
reliable technique for inferring the physical and dynamical state of gas in the
intervening absorption clouds at high redshifts, $z \ga 2$. Of particular interest
are the measurements of metallic absorptions in the optically thin diffuse clouds
with neutral hydrogen column densities $N$(\ion{H}{i}) $\la 3\times10^{17}$ cm$^{-2}$.
These systems will be called as `Ly-$\alpha$ absorbers' (LAA) 
to distinguish from the
damped Ly-$\alpha$ absorption systems (DLA) which show much higher column densities,
$N$(\ion{H}{i})~$\ga 2\times10^{20}$ cm$^{-2}$. The latter are believed 
to arise in the galactic disks (e.g., Wolfe et al. 1995), whereas the former may be
physically related to the external ($\sim$ 10-100 kpc-scale) regions of galaxies
(e.g., Chen et al. 2001). Thus the measurements of the LAAs can provide fundamental
insights into conditions prevailing in the galactic environments (external halos)
in the early universe.

Since these regions are mainly
photoionized by the local metagalactic UV radiation,
the ionization states of the LAAs are sensitive 
to the spectral shape of the background
radiation in the 1-5 Ryd range. The spectral energy distribution in the metagalactic
ionizing background is defined in turn by the QSO continua  
filtered through the quasar environments and the IGM. 
Two implications of the LAA analysis  -- the metal content
of the external halos and the spectral
shape of the local UV background -- are, therefore, tightly coupled. 

To clarify 
the mechanism of the metal enrichment, accurate measurements of the metal
abundances in the LAAs are required. The main problem here is how to 
account for the ionization correction. In general, 
the contribution to the line intensity $I_\lambda$  
within the profile comes from all volume elements 
distributed along the line of sight and having the same radial velocity.
If the gas number density, $n_{\rm H}$, varies from point to point, then
the intensity $I_\lambda$ is caused by a
{\it superposition} of different ionization states.

Recently, we have developed a method called `Monte Carlo inversion'
(MCI) to recover the physical
parameters of the LAAs assuming that the absorbing cloud is a continuous region with
fluctuating density and velocity fields 
(Levshakov et al. 2000, hereafter LAK).
The MCI was applied to high quality QSO spectra obtained with the
VLT/UVES, Keck/HIRES, and HST/STIS (Levshakov et al. 2002; 
Levshakov et al. 2003a,b,c,d,e). The results of these studies are briefly reviewed
in this contribution.

\section{The MCI procedure}

The layout of the MCI procedure  is the following.
We assume that the metal abundances within the absorber
are constant, the gas is optically thin for the ionizing UV radiation, and the gas
is in thermal and ionization equilibrium. 
The radial velocity  $v(x)$ and
the total hydrogen density $n_{\rm H}(x)$
along the line of sight are considered as two random fields
which are represented by their sampled values
at equally spaced intervals $\Delta x$,
i.e. by the vectors
$\{ v_1, \ldots, v_k \}$ and $\{ n_1, \ldots, n_k \}$
with $k$ large enough ($\sim 150-200$)
to describe the narrowest components of the complex spectral lines.
The radial velocity is assumed to be normal distributed with the
dispersion $\sigma_v$, whereas the gas density is 
log-normal distributed with the mean $n_0$
and the dispersion $\sigma_y$
($y = n_{\rm H}/n_0$).
The stochastic fields are approximated by
the Markovian processes.
The accuracy of the restoring procedure depends on 
the number of different ions in different ionization
stages involved in the analysis of a given LAA.   

A set of the fitting parameters 
in the least-squares minimization  of the objective function
(Eqs.~[29] and [30] in LAK) includes
$\sigma_v$ and $\sigma_y$ along with
the total hydrogen column density $N_{\rm H}$,
the mean ionization parameter $U_0$,
and the metal abundances $Z_a$
for $a$ elements observed in the LAA.
With these parameters we can further calculate
the mean gas number density $n_0$,
the column densities for different species $N_{\rm a}$,
the mean kinetic temperature  
$T_{\rm kin}$, and the line-of-sight size $L = N_{\rm H}/n_0$ of the absorber
(note that $n_0$ and, correspondingly, $L$ scales with the intensity
of the local background radiation field).

We start calculations assuming some standard 
ionizing spectrum [e.g., power law, Mathews \& Ferland (1987), or
Haardt \& Madau (1996, hereafter HM)] and
compute the fractional ionizations and the kinetic temperatures at each point
$x$ along the sightline 
with the photoionization code CLOUDY (Ferland 1997).
To optimize the patterns of $\{v_i\}$ and $\{n_i\}$ and to estimate
simultaneously the fitting parameters, 
the simulated annealing algorithm with Tsallis
acceptance rule and an adaptive annealing temperature choice
are used.
If the fitting with the standard spectrum is impossible,
its shape is adjusted using the procedure based on 
the experimental design technique (Levshakov et al. 2003e).  

\begin{figure}[t]
\plotone{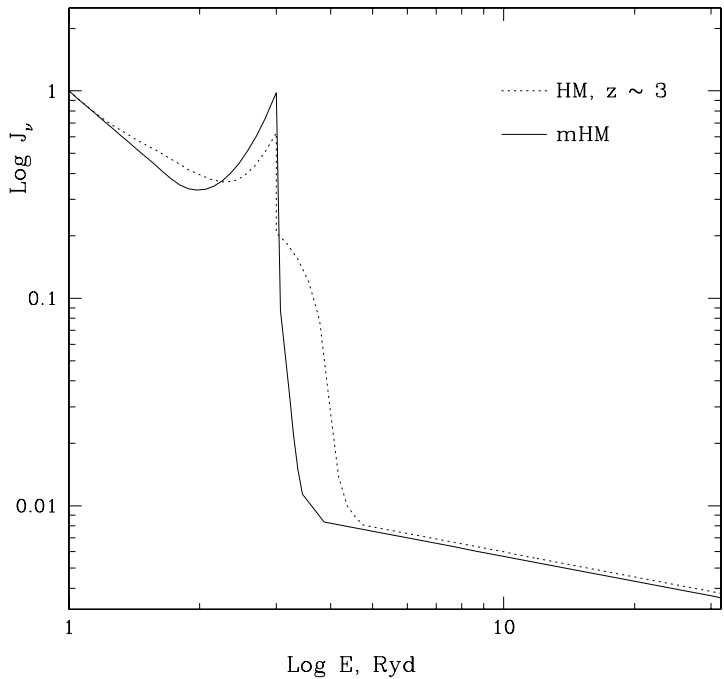}
\caption{A typical metagalactic
ionizing spectrum at redshift $z~\sim~3$ (the dotted curve)
modeled by Haardt \& Madau (1996),
and its modification (the solid curve) 
required to match the absorption lines observed in a LAA 
at $z = 2.82$ toward HE~0940--1050 (Levshakov et al. 2003e).
The spectrum is normalized so that $J_\nu(h\nu =$ 1 Ryd) = 1.
The emission bump at 3 Ryd is caused by reemission of \ion{He}{ii} Ly$\alpha$
and two-photon continuum emission
from intergalactic clouds. }
\end{figure}

\begin{figure}[t]
\plotone{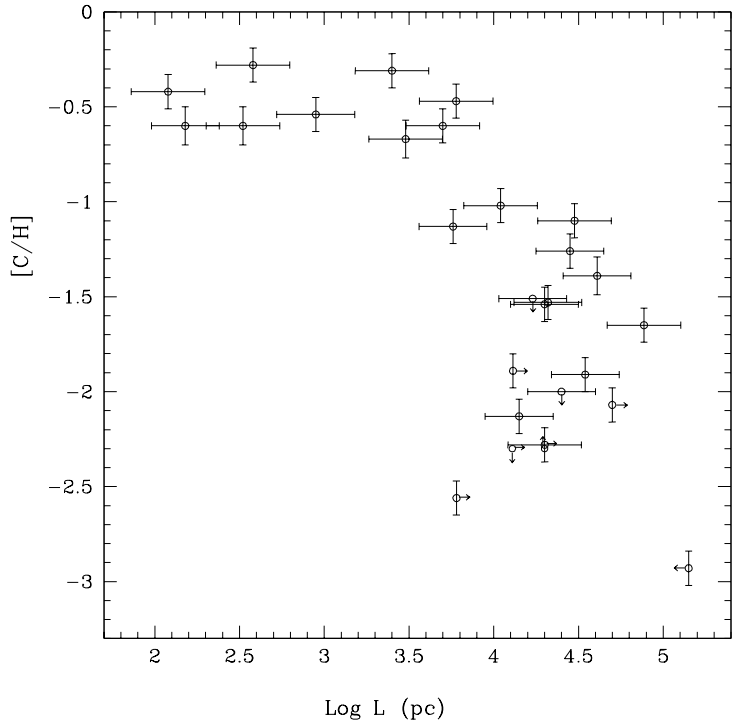}
\vspace{0.3cm}
\caption{Carbon abundances [C/H]
plotted against the logarithmic linear
size $L$  of the absorber estimated by the 
MCI procedure (Levshakov et al. 2002; 2003a,b,c,e).
[C/H] decreases with increasing $L$ 
reflecting, probably, the dilution of metals within galactic halos
caused by the mass transport processes.
}
\end{figure}

\begin{figure}[t]
\plotone{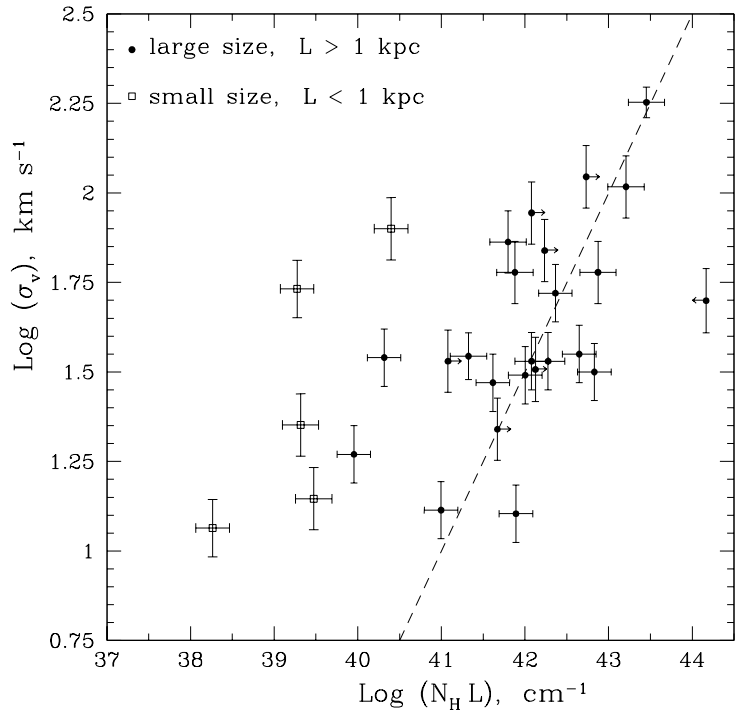}
\caption{Plot of the line of sight velocity
dispersion $\log (\sigma_v)$ vs.  $\log (N_{\rm H}\,L)$ for
the same sample of the LAAs shown in Fig.~2. 
The dashed line corresponds to the relation $\log (\sigma_v) \propto
0.5\log (N_{\rm H}\,L)$ expected for the virialized systems.
Open squares represent HVC-like clouds with $L \la 1$ kpc.
}
\end{figure}

\section{The spectral shape of the ionizing continuum}

In practice, the shape of the ionizing spectrum can 
be estimated in cases when the value of
$N$(\ion{H}{i}) is measured accurately and the absorption 
system contains unsaturated lines
of at least \ion{C}{ii}--\ion{C}{iv} and \ion{Si}{ii}--\ion{Si}{iv}, otherwise
the system is not sensitive to the fine tuning of the continuum shape.

We estimated the spectral shape for the LAAs at
$z = 2.82$
(\ion{H}{i}, \ion{C}{ii}, \ion{C}{iii}, \ion{Si}{iii}, \ion{C}{iv}, \ion{Si}{iv})
toward HE~0940--1050;
$z = 2.7711$
(\ion{H}{i}, \ion{C}{ii}, \ion{Si}{ii}, \ion{C}{iii}, \ion{N}{iii},
\ion{Si}{iii}, \ion{C}{iv}, \ion{Si}{iv}, \ion{O}{vi}), 
and $z = 2.94$
(\ion{H}{i}, \ion{C}{ii}, \ion{Si}{ii}, \ion{C}{iii}, 
\ion{Si}{iii}, \ion{C}{iv}, \ion{Si}{iv})
toward Q~1157+3143 (Levshakov et al. 2003e).

All three recovered spectra of ionizing radiation show common features:
a bump at $E = 3$ Ryd, which is more pronounced comparing to the model
mean intergalactic UV spectra at $z = 3$
like that of Haardt \& Madau (1996),
and a sharp break just after the bump -- also
at variance with the model predicting a smeared out break at $E = 4$ Ryd due
to ionization of \ion{He}{ii}. 
An example of ionizing background estimated for the $z = 2.82$ absorption system
toward HE~0940--1050 is 
shown in Fig.~1.
This spectral shape rules out
a considerable galactic
contribution to the QSO dominated UV ionizing background  at $z \sim 3$.
The recovered UV spectrum can be well explained in the scenario
of the delayed re-ionization of \ion{He}{ii} (Reimers et al. 1997).
In this case the sharp break at $E = 3$ Ryd
occurs due to strong resonant scattering of QSO radiation
in metal and \ion{He}{ii}
Lyman series lines whereas a part of the absorbed photons re-emitted
by the intergalactic gas in
the \ion{He}{ii} Ly$\alpha$
and two-photon continuum emission increases
the amplitude of the bump at 3 Ryd.
Our results also indicate that the re-ionization of \ion{He}{ii}
has not been
yet completed by $z = 2.77$.
This results is in line with recent observations of the \ion{He}{ii} Ly-$\alpha$
forest by Shull et al. (2003) who claim that `the ionizing background is highly
variable throughout the IGM' at $z \sim 2.8$.

\section{`[C/H]$-L$' and `$\sigma_v-N_{\rm H}L$' relations}

The analyzed LAAs show that they are a {\it heterogeneous} population
that is formed by at least three groups of absorbers~:
(1) extended metal-poor ($Z < 0.1\,Z_\odot$) gas halos of distant galaxies;
(2) gas in dwarf galaxies ($0.1Z_\odot < Z \la 0.3\,Z_\odot$); and
(3) metal-enriched gas ($Z \ga 0.5\,Z_\odot$) arising from the inner
galactic regions and condensing into the clouds within
the hot galactic halo (high redshift analogs to the Galactic 
high velocity clouds, HVC).

Figure~2  shows a plot of the measured carbon
abundances [C/H]\footnote{Using the customary definition
[X/H] = log\,(X/H) -- log\,(X/H)$_\odot$. Photospheric solar abundances
are taken from Holweger (2001).}
versus logarithmic sizes of the studied systems.
Systematically higher abundances are seen in compact systems.
This tendency reflects,
probably, the dilution of metals within galactic halos
caused by the mass transport processes (like diffusion, turbulent mixing,
galactic rotation, shear, and etc.).
Metals are, probably, 
transported into the halo in form of small dense clouds
carried out by wind or jets. In some cases we directly observe such blobs.
For instance, the LAAs at $z = 1.385$ ([C/H] $\simeq -0.3$, $L \simeq 1.7-2.5$ kpc)
and at $z = 1.667$ ([C/H] $\simeq -0.5$, $L \simeq 1$ kpc) toward HE~0515--4414
as well as that at $z = 2.966$ ([C/H] $\simeq -0.4$, $L \simeq 100$ pc) toward
Q~0347--3819
are embedded in extremely metal-poor halos with [C/H]~$< -2$ 
(Levshakov et al. 2003b,c). 

If LAAs are formed in gas clouds gravitationally bound with
intervening galaxies, their internal kinematics should be
closely related to the total masses of the host galaxies. 
We find a correlation between
the absorber's linear size $L$ and its
line-of-sight velocity dispersion $\sigma_v$.
The virial theorem  states~:
$\sigma^2_v \propto {M}/{L} \propto n_0\,L^2 = N_{\rm H}\,L$.
Assuming that the gas systems are in quasi-equilibrium,
one can expect $\sigma_v \propto (N_{\rm H}\,L)^{1/2}$.
In Fig.~3 we plot the measured values of 
$\sigma_v$ versus the product of $L$ and 
the total gas column density $N_{\rm H}$.
It is seen that most
systems with linear sizes $L > 1$ kpc lie along the line with the slope 0.5.
Taking into account that we know neither the impact parameters nor the halo
density distributions, this result can be considered as a quite good fit to
the expected relation for the virialized systems.
Hence we may conclude that most absorbers with $L > 1$ kpc are
gravitationally bound with systems that appear to be in virial
equilibrium at the cosmic time when the corresponding LAAs
were formed.

\acknowledgments{This work would not have been possible without the
contribution of many individuals, including Irina Agafonova, Wilhelm Kegel,
Miriam Centuri\'on, Paolo Molaro, Igor Mazets, Miroslava Dessauges-Zavadsky,
Sandro D'Odorico, Dieter Reimers, Robert Baade, David Tytler, and Art Wolfe.
I also appreciate support from the RFBR grant No. 03-02-17522, and I am very
grateful to the IAU for a travel grant.}

\end{document}